\documentclass[%
 aip,
 amsmath,amssymb,
 reprint,%
]{revtex4-1}

\usepackage{graphicx}% Include figure files
\usepackage{dcolumn}% Align table columns on decimal point
\usepackage{bm}% bold math
\usepackage[utf8]{inputenc}
\usepackage[T1]{fontenc}
\usepackage{mathptmx}
\usepackage{etoolbox}

\makeatletter
\def\@email#1#2{%
 \endgroup
 \patchcmd{\titleblock@produce}
  {\frontmatter@RRAPformat}
  {\frontmatter@RRAPformat{\produce@RRAP{*#1\href{mailto:#2}{#2}}}\frontmatter@RRAPformat}
  {}{}
}%
\makeatother
\begin{document}

\preprint{AIP/123-QED}

\title[Metasurface Tape for Efficient Millimeter-Wave Power Transfer]{Metasurface Tape for Efficient Millimeter-Wave Power Transfer via Surface-Wave Propagation}

% Force line breaks with \\
\author{Phuc Toan Dang}
\author{Kota Suzuki}
\affiliation{Department of Engineering, Graduate School of Engineering, Nagoya Institute of Technology, Gokiso-cho, Showa, Nagoya, Aichi, 466-8555, Japan%\\This line break forced with \textbackslash\textbackslash
}%
\author{Yoshiki Ashikaga}
\author{Yasushi Tsuchiya}
\affiliation{Teraoka Seisakusyo Co., Ltd., 1-4-22 Hiromachi, Shinagawa, Tokyo, 140-8711, Japan%\\This line break forced with \textbackslash\textbackslash
}%
\author{Sendy Phang}%
\affiliation{George Green Institute for Electromagnetics Research, Faculty of Engineering, University of Nottingham, University Park, Nottingham NG7 2RD, United Kingdom%\\This line break forced with \textbackslash\textbackslash
}%
\author{Hiroki Wakatsuchi}
 \email{wakatsuchi.hiroki@nitech.ac.jp.}
\affiliation{Department of Engineering, Graduate School of Engineering, Nagoya Institute of Technology, Gokiso-cho, Showa, Nagoya, Aichi, 466-8555, Japan%\\This line break forced with \textbackslash\textbackslash
}%

\date{\today}% It is always \today, today,

\begin{abstract}
Millimeter-wave technologies are essential for future high-speed wireless communications. However, a fundamental challenge remains in the form of severe free-space path loss, where the power density decreases inversely with the square of the distance $r$ (i.e., $\propto r^{-2}$) as a spherical dependence. To overcome this limitation, we propose a flexible metasurface tape that is designed to guide electromagnetic energy as surface waves. Unlike conventional free-space propagation, this engineered metasurface confines the field to a subwavelength interface, thereby altering the power decay law to a circular dependence (i.e., $\propto r^{-1}$). We numerically and experimentally, for the first time, demonstrate this concept using a periodic grounded-patch array fabricated on a flexible substrate and operated at approximately 100 GHz. The measurement results show that the metasurface tape significantly increases the transmitted power, yielding an average rate of improvement of approximately 40 per meter in received power relative to the free-space baseline in our measurement geometry (e.g., 29-dB increase at 2 m). This increase is realized over a broad bandwidth from 95 GHz to 105 GHz (i.e., approximately 10 \%), accommodating wideband modulation schemes required for high-data-rate applications. The flexible, lightweight nature of the tape allows it to be easily installed on diverse surfaces. Our demonstration indicates that the metasurface tape is a promising platform for extending the effective range of millimeter-wave systems, thus offering a robust solution to the path-loss bottleneck in next-generation wireless networks.
\end{abstract}

\maketitle

Metasurfaces (MSs) are planar analogs of metamaterials\cite{smithDNG1D, smithDNG2D2, MTMbookEngheta, calozBook} that consist of ultrathin arrangements of subwavelength resonant elements.\cite{EBGdevelopment, yu2011light, yu2014flat} Owing to their low profiles, lightness, and fabrication compatibility, MSs have emerged as practical platforms for electromagnetic-wave engineering over a broad spectral range from microwaves to the visible spectrum.\cite{sievenpiper2003two, pfeiffer2013metamaterial, yu2011light} By adjusting the geometry and orientation of the constituent resonators, MSs realize spatially tailored boundary conditions that enable deterministic control of wave behavior, including reflection/transmission characteristics and the local phase, amplitude, polarization state, and effective impedance. Whereas early studies primarily emphasized free-space wavefront manipulation,\cite{pfeiffer2013metamaterial, yu2011light} recent advances have increasingly focused on the control of guided modes and on engineered coupling between bound surface waves and radiating fields.\cite{sun2012gradient, li2017nonlinear, li2018metasurfaces, suzuki2024improved} Because surface waves are tightly confined to an interface and decay rapidly away from it, judiciously designed MSs provide versatile approaches for steering, focusing, and routing energy along their surfaces.\cite{sievenpiper2010, luo2015self, lee2016patterning, suzuki2024experimental} These characteristics are particularly attractive for wireless communications and wireless power transfer (WPT), where MS-assisted propagation channels can improve transfer efficiency and extend the operational range.\cite{di2020smart, gradoni1smart}

\begin{figure}[tb!]
\includegraphics[width=0.95\linewidth]{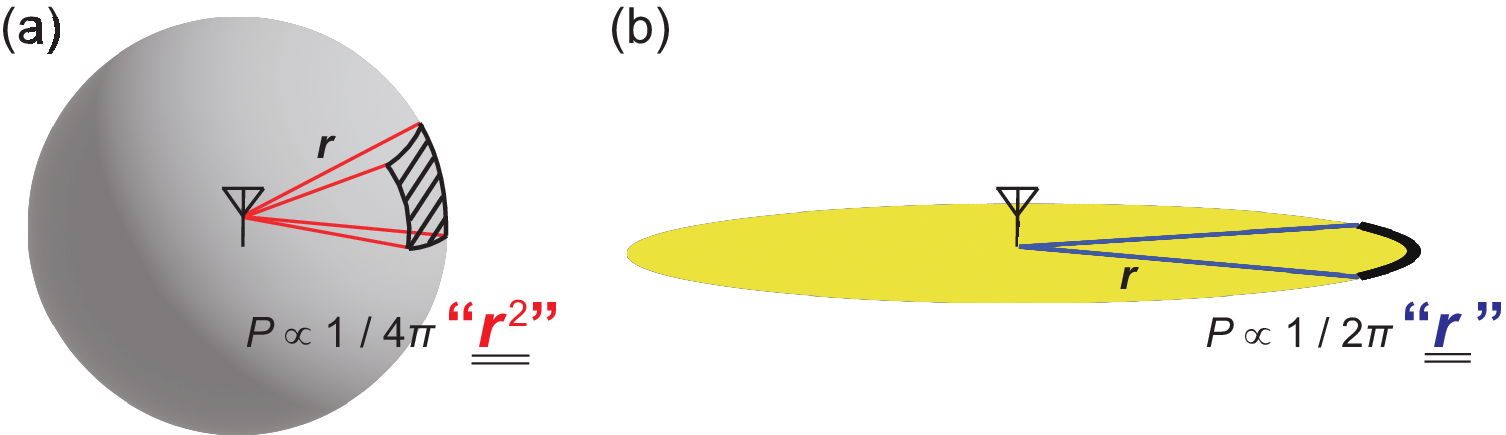}% Here is how to import EPS art
\caption{\label{fig:1} Energy spreading. (a) Conventional free-space propagation approach. (b) Proposed surface-wave propagation approach. }
\end{figure}

A central limitation in conventional WPT that relies on free-space propagation is spatial energy spreading: the power flux from an isotropic radiator decreases with the surface area of a sphere, scaling as $P/(4\pi r^2)$,\cite{balanis2005antenna} where $P$ and $r$ denote the radiated power and the distance to the observation point, respectively (see Fig.\ \ref{fig:1}a). Moreover, the Friis transmission equation indicates that the path loss becomes increasingly severe at higher carrier frequencies. In contrast, an MS can redirect radiation energy into a surface-wave channel, thereby confining the energy to the interface such that, in an idealized, lossless scenario, the power is distributed over a circular perimeter and scales more weakly as $P/(2\pi r)$ (Fig.\ \ref{fig:1}b). This reduced spreading suggests a promising strategy for increasing long-distance power-transfer efficiency, provided that dissipative and radiative losses are properly managed. In particular, such surface-wave-guiding MSs provide a powerful solution for maintaining the energy of strongly attenuating millimeter-wave signals for next-generation wireless communications. Implementing conducting patterns on flexible substrates\cite{walia2015flexible} rather than on rigid printed circuit boards (PCBs) would greatly improve conformability and enable deployment on diverse surfaces in practical environments.

Here, we propose a flexible MS tape that enables efficient millimeter-wave power transfer via surface-wave propagation. Unlike other structures,\cite{MTMbookEngheta, calozBook, yu2014flat, takeshita2024frequency} our MSs do not rely on narrow-band resonant mechanisms to support surface waves, thereby enabling efficient power transfer over a broader frequency range. Additionally, the use of a flexible tape substrate ensures that our MSs can be readily installed on diverse surfaces and materials. Our results demonstrate that, compared with free-space propagation, the transmitted energy is significantly increased by a factor of approximately 40 per meter at 100 GHz (e.g., 29-dB increase at 2 m). This outstanding performance is observed from 95 GHz to 105 GHz, thereby supporting advanced wideband modulation techniques.

\begin{figure}[tb!]
\includegraphics[width=0.95\linewidth]{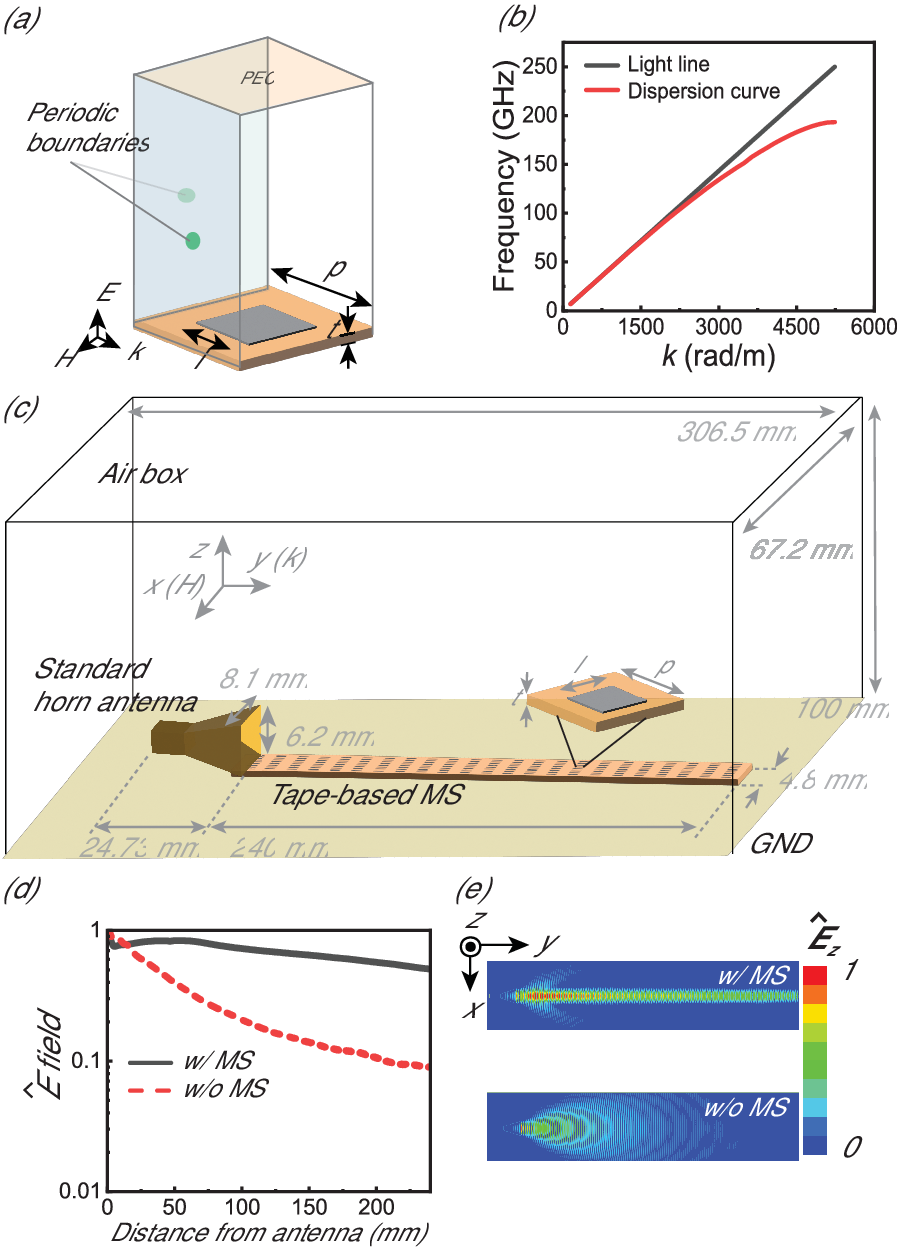}% Here is how to import EPS art
\caption{\label{fig:2} Design of the MS tape via numerical simulations. (a) Design of the periodic unit cell. (b) Dispersion diagram. (c) Full-wave simulation model. (d) Normalized $E$ field ($\hat{E}$) as a function of the propagation distance at 100 GHz. (e) Simulated field distributions with (top) and without (bottom) the MS. }
\end{figure}

We first designed an MS unit cell that supports surface waves, as shown in Fig.\ \ref{fig:2}a. The unit cell consists of a periodic square patch array (with edge length $l=$ 0.3 mm and periodicity $p=$ 0.6 mm) printed on dielectric tape (with thickness $t=$ 0.1 mm and relative permittivity $\epsilon_r=$ 2.458) backed by a ground (GND) plane. Such grounded patch arrays support a transverse magnetic (TM) wave as a fundamental mode.\cite{EBGdevelopment, quarfoth2013artificial} Using the eigenmode solver in ANSYS Electronics Desktop (2023 R2),\cite{suzuki2024improved} we obtained the dispersion diagram shown in Fig.\ \ref{fig:2}b. In this diagram, the dispersion curve closely follows the light line up to approximately 100 GHz, indicating that the fundamental surface-wave mode is supported in this frequency range. Above approximately 100 GHz, the MS begins to behave as a high-impedance surface and no longer supports the propagation of the fundamental mode. Accordingly, we mostly focused on the 95--105 GHz band, where the mode remains well supported, and we additionally included 110 GHz to illustrate the performance degradation as the structure approaches the high-impedance region.

Next, we designed a long MS tape model and compared its performance in power transfer with that without the MS as a typical approach for free-space propagation. The MS tape model was composed of 8 $\times$ 400 unit cells (4.8 mm wide and 240 mm long) and deployed in front of a standard horn antenna (compatible with WR-8) above a GND plane, as shown in Fig.\ \ref{fig:2}c. Under this condition, as shown in Fig.\ \ref{fig:2}d, the horn antenna radiated an electromagnetic wave at 100 GHz that became strongly attenuated without the MS in free space, which was predicted earlier with the Friis transmission equation. In contrast, the radiated energy was maintained at almost the same power when the MS tape was introduced. This substantial difference is clarified in Fig.\ \ref{fig:2}e, where the electric field ($E$-field) distributions with and without the MS are visualized. This figure shows that the radiated $E$ field was strongly concentrated along the MS, which resulted in efficient wave propagation over a long distance of 80 wavelengths even at a high frequency of 100 GHz. This feature clearly contrasts with the results obtained without MS tape, where the radiated energy was widely distributed throughout three-dimensional space. Thus, these results indicate that the proposed MS successfully achieved efficient power transfer of a millimeter wave by guiding the energy along the MS tape substrate.

\begin{figure}[tb!]
\includegraphics[width=0.95\linewidth]{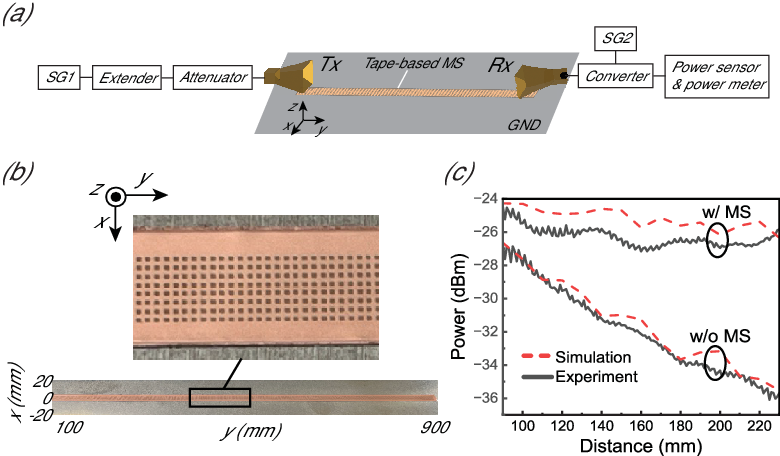}% Here is how to import EPS art
\caption{\label{fig:3} Experimental demonstration. (a) Measurement setup. The Rx position was changed along the MS tape. (b) Prototype of the MS tape. (c) Measured power profiles at 100 GHz and corresponding simulation results. }
\end{figure}

The proposed MS was then fabricated and experimentally evaluated using the measurement setup shown in Fig.\ \ref{fig:3}a (see also the actual measurement setup in Fig.\ \ref{figS:1} of the Supplemental Material). In this setup, a tape-based MS prototype was tested at approximately 100 GHz, which was achieved by frequency-multiplying a lower-frequency source signal.\cite{takimoto2025millimeter} Specifically, a signal generator (SG) (Anritsu, MG3692C) generated an input signal at 11.1 GHz, which was up-converted to 99.9 GHz by using a frequency extender (Virginia Diodes, Inc., VDI-730SGX). The frequency-extended signal was radiated from a standard horn antenna (Pasternack, PEWAN1022) and guided as a surface wave along the MS tape. This energy was observed by another standard horn antenna connected to a down converter (Virginia Diodes, Inc., VDI-731SAX) via a coaxial cable and an adapter. Therefore, while the MS supported surface-wave propagation at 99.9 GHz (or approximately 100 GHz), the received signal was converted to 3.9 GHz to be measured by a power sensor (Anritsu, MA2473D) connected to a power meter (Anritsu, ML2496A). The downconverted frequency was determined by an additional SG (Keysight Technologies, N5183B) that provided an 8-GHz local oscillator (LO) signal. In this measurement, the receiver location was varied in 1-mm steps via a two-dimensional (2D) scanning system to obtain not only one-dimensional (1D) power profiles but also 2D power distributions with and without the MS tape. The actual tape-based MS prototype is shown in Fig.\ \ref{fig:3}b, where the sample was deployed along the incident wave direction (the $y$-axis direction) on an aluminum plate. The sample was fabricated as designed in the simulation model in Fig.\ \ref{fig:2}. Periodic conducting patches were printed using silver ink on an adhesive acrylic substrate with the same physical dimensions as those in the simulation model. For this setup, the power measured by the power meter via the second horn antenna is shown in Fig.\ \ref{fig:3}c, which experimentally validates the ability to receive a higher-power signal with the MS tape even at a location farther than possible without the MS tape. Additionally, these results closely agreed with the simulation results. Notably, these results differ slightly from those in Fig.\ \ref{fig:2} because the simulation model in Fig.\ \ref{fig:3}c includes a receiver antenna to achieve consistency with the measurement setup in Fig.\ \ref{fig:3}a. In contrast, the simulation model in Fig.\ \ref{fig:2}c consisted only of a transmitting antenna. This difference produced minor standing waves in the simulation results shown in Fig.\ \ref{fig:3}c. The same trend was observed in the corresponding measurement results, which had relatively limited fluctuations, as the receiver was surrounded by absorbing foam during the measurement. Despite this minor difference, the measurement results shown in Fig.\ \ref{fig:3}c confirm that compared with free-space propagation (i.e., without the MS tape), the proposed MS tape strongly guided an incident, high-frequency signal as a surface wave.

\begin{figure}[tb!]
\includegraphics[width=0.95\linewidth]{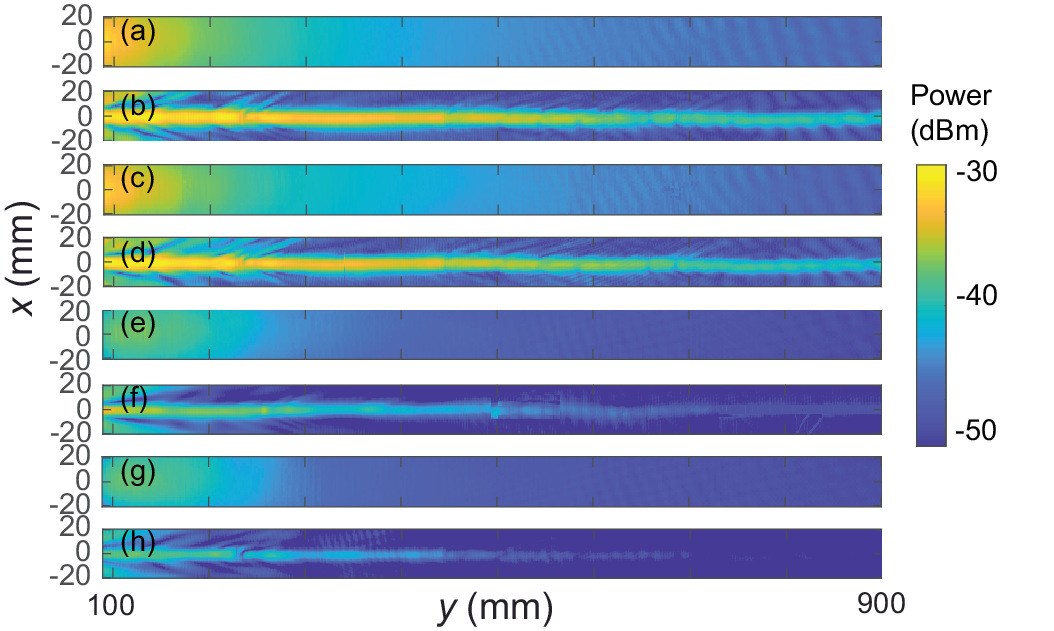}% Here is how to import EPS art
\caption{\label{fig:4} Experimentally observed 2D power distributions at various frequencies. Results are shown at (a, b) 95 GHz, (c, d) 100 GHz, (e, f) 105 GHz, and (g, h) 110 GHz (a, c, e, g) without and (b, d, f, h) with the MS tape. }
\end{figure}

We next measured the 2D profiles for the MS tape at various frequencies, as shown in Fig.\ \ref{fig:4}. This figure shows that without the MS tape, the waves generated from the transmitting antenna propagated as free-space waves and quickly became attenuated at any frequency between 95 GHz and 110 GHz. However, our MS tape, which was fabricated from a low-cost tape medium with a simple periodic pattern, strongly supported the incident waves, thereby enabling the surface waves to propagate farther with minimal attenuation.

\begin{figure}[tb!]
\includegraphics[width=0.95\linewidth]{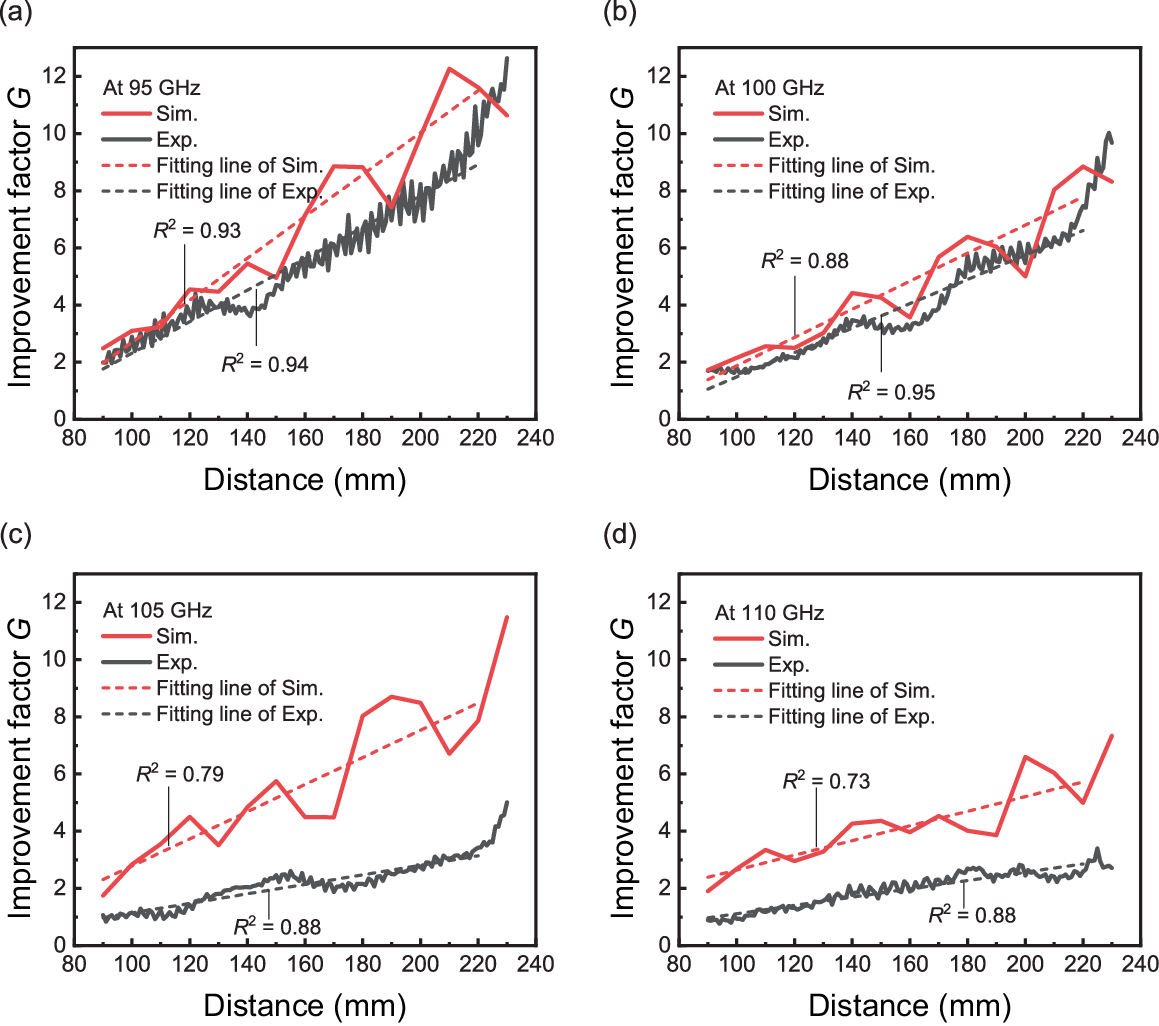}% Here is how to import EPS art
\caption{\label{fig:5} Improvement factor $G$ in millimeter-wave power transfer. Results are shown at (a) 95 GHz, (b) 100 GHz, (c) 105 GHz, and (d) 110 GHz. The coefficients of determination of the linear fitting lines are represented by $R^2$. The fitting lines are obtained between 90 mm and 220 mm.}
\end{figure}

To evaluate the performance quantitatively, the gap between the two cases (i.e., with and without the MS tape) is more clearly shown in Fig.\ \ref{fig:5}. To quantify the relative benefit of the MS tape, we define the improvement factor $G$ as
\begin{eqnarray}
    G(r)=\frac{P_{MS}(r)}{P_{FS}(r)},
\end{eqnarray}
where $P_{MS}$ and $P_{FS}$ are the received powers expressed in watts. In an ideal lossless scenario, $P_{FS}$ is proportional to $r^{-2}$ for free-space propagation, whereas a radially spreading surface wave exhibits a weaker $r^{-1}$ dependence; hence $G(r)$ is proportional to $r$. Following this assumption, the received power was increased in proportion to the propagation distance at any frequency between 95 and 110 GHz (see the linear fitting curves in Fig.\ \ref{fig:5}). In particular, relatively large increases were experimentally observed at 95 and 100 GHz, whereas the improvement at 110 GHz was comparatively small, which is attributed to the frequency approaching the high-impedance region, as predicted by the eigenmode simulation in Fig.\ \ref{fig:2}b (see the dispersion curve diverging from the light line at approximately 110 GHz). Additionally, these experimental results were also in close agreement with the numerically derived results. Notably, these simulation results were more strongly influenced by the standing wave issue than the measurement results were, as explained in Fig.\ \ref{fig:3}. Despite this issue, both the measurement and the simulation results showed that the received power was markedly increased in a broad frequency range between 95 and 110 GHz when our MS tape was introduced.

Moreover, to derive a generalized index independent of the tape length, we estimated the power improvement per unit length, as shown in Fig.\ \ref{fig:6} (see also Fig.\ \ref{figS:2} in the Supplemental Material for the phase change). The rate of improvement was calculated by using the gradients of the fitting functions shown in Fig.\ \ref{fig:5}. However, owing to the presence of the standing waves in Fig.\ \ref{fig:5}, the fitting curves (and their gradients) could vary depending on the fitting distance range  (see Fig.\ \ref{figS:3} and Fig.\ \ref{figS:4} in the Supplemental Material for the setting of fitting functions and its influence on the rate of improvement, respectively). Thus, such variations are denoted by error bars that represent standard deviations in Fig.\ \ref{fig:6}. Despite these variations, the simulation and measurement results exhibited similar trends. First, both types of result showed that the received power was significantly improved by a factor of at least 10 per meter from 95 to 110 GHz, indicating that the proposed MS-tape approach is not only experimentally feasible but also an effective solution for transmitting high-frequency signals to a distant location. For instance, according to the simulation results, the received power with the MS tape becomes approximately 100 to 150 times greater than that without the MS tape at 2 meters from 95 to 105 GHz, thus ensuring a bandwidth of 10 \% with a center frequency of 100 GHz. Notably, although the operating frequency is different, this fractional bandwidth ($\sim$10 \%) is sufficiently wider than that applied to many modern modulation schemes, such as the orthogonal frequency division multiplexing (OFDM) adopted in Wi-Fi in the 5-GHz band.\cite{goldsmith2005wireless} In both the simulations and the measurements, the rate of improvement decreased with increasing frequency. This reduction can be explained by the presence of a high-impedance region above 110 GHz (again see Fig.\ \ref{fig:2}). Here, we stress that in the simulations, the rate of improvement was relatively small at 100 GHz because the simulation model was much larger than the wavelength of the incident wave, which made it difficult to mesh the entire model in exactly the same manner throughout all these simulations. Therefore, the difference in the meshing precision potentially led to the variation in the computational accuracy. Additionally, the experimental performance was slightly worse than the numerical performance, which was presumably because of an additional loss in the tape medium. The printing accuracy could influence the measurement sample by reducing the actual conducting patch dimensions and shifting the overall frequency response toward lower frequencies. Nonetheless, our results shown in Fig.\ \ref{fig:2} to Fig.\ \ref{fig:6} demonstrate that the proposed design concept can transfer the energy of millimeter waves much more efficiently than conventional free-space wireless power transfer approaches can over a wide range of frequencies, thereby ensuring full support of wideband modulated signals. Thus, our MS tape not only achieves fundamentally improved transmission performance but also provides an engineering solution that enables next-generation wireless communications using the millimeter-wave frequency band. 

\begin{figure}[tb!]
\includegraphics[width=0.95\linewidth]{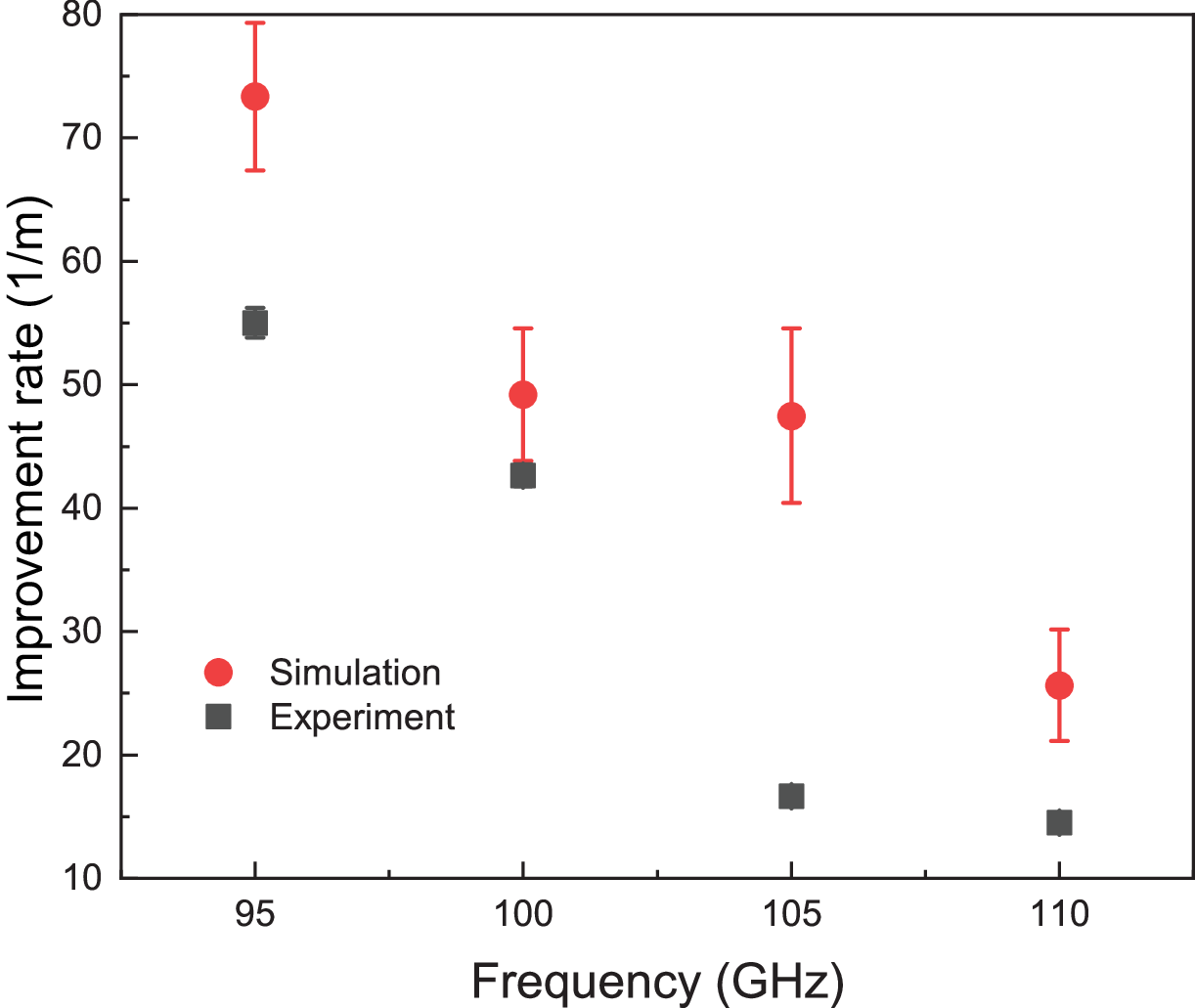}% Here is how to import EPS art
\caption{\label{fig:6} Rate of improvement factor $G$ in millimeter-wave power transfer as a function of frequency. }
\end{figure}

In conclusion, we developed a flexible MS tape that can significantly increase the millimeter-wave power transfer efficiency via surface-wave propagation. By confining electromagnetic fields to the MS interface, we successfully transformed the propagation channel from a three-dimensional spherical decay ($\propto r^{-2}$) to a quasi-two-dimensional cylindrical decay ($\propto r^{-1}$). Experimental demonstrations at 100 GHz verified that the prototype improved the received power by a factor of approximately 40 per meter compared with free-space transmission. Furthermore, the proposed structure maintains high performance across a wide bandwidth from 95 GHz to 105 GHz, thus offering sufficient spectral resources for advanced modulation schemes. Given its low profile, flexible form factor, and ability to mitigate severe path loss, this MS tape provides a practical solution for extending the range of next-generation wireless communication and power transfer systems.

This study was supported by the Japan Science and Technology Agency (JST) under Adopting Sustainable Partnerships for Innovative Research Ecosystem (ASPIRE) (No.\ JPMJAP2431), KAKENHI grant from the Japan Society for the Promotion of Science (JSPS) (No.\ 24KF0152), and the Ministry of Internal Affairs and Communications (MIC) under the Fundamental Technologies for Sustainable Efficient Radio Wave Use R\&D Project (FORWARD) (No.\ JPMI250610002).

\subsection*{AUTHOR DECLARATIONS\\Conflict of Interest}
\vspace{-5mm}The authors have no conflicts to disclose.

\subsection*{DATA AVAILABILITY}
\vspace{-5mm}The data that support the findings of this study are available from the corresponding author upon reasonable request.

\subsection*{REFERENCES}\vspace{-5mm}

%\nocite{*}
%\bibliography{aipsamp}% Produces the bibliography via BibTeX.
%\bibliography{cwEqCircuits}% Produces the bibliography via BibTeX.

\providecommand{\noopsort}[1]{}\providecommand{\singleletter}[1]{#1}%

\clearpage

%%%%%%%%%%%%%%%%%%%%%%%%%%%%%%%%%%%%%%%%%%%%%%%%%%%%
\renewcommand{\appendixname}{Supplemental Material}
%\pagebreak
%\widetext
\begin{center}
{\Large \hspace{-4mm}\textit{Supplemental Material for}}\vspace{3mm}\\
\textbf{Metasurface Tape for Efficient Millimeter-Wave Power Transfer via Surface-Wave Propagation}\vspace{3mm}\\
\text{Phuc Toan Dang, Kota Suzuki, Yoshiki Ashikaga,}\\
\text{Yasushi Tsuchiya, Sendy Phang, and Hiroki Wakatsuchi}\\
\end{center}

\renewcommand{\thesection}{\arabic{section}}
%%%%%%%%%%%%%%%%--------------%%%%%%%%
\appendix
%\section{The first thing in SI
 %\label{sec:si1}}
%another word
%\subsection{The first sub-thing in SI}
%bla bla
%\subsection{The second sub-thing in SI}
%bla bla 2
%\section*{Supplementary Material}
\renewcommand{\thefigure}{S\arabic{figure}}
\setcounter{figure}{0}
\renewcommand{\thetable}{S\arabic{table}}
\setcounter{table}{0}
\setcounter{page}{1}
\renewcommand{\thepage}{S\arabic{page}}

\begin{figure}[b!]
\includegraphics[width=0.95\linewidth]{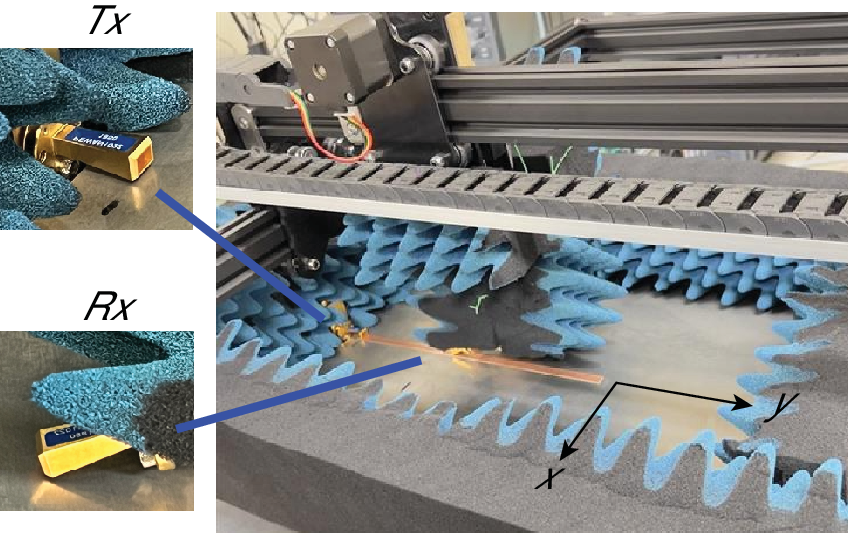}% Here is how to import EPS art
\caption{\label{figS:1} The actual measurement setup. }
\end{figure}

The measurement results reported in Figs.\ \ref{fig:3} to \ref{fig:6} were obtained by using the measurement setup shown in Fig.\ \ref{figS:1}. In this setup, we introduced a scanning system to move the receiving antenna and visualize 2D field distributions with and without our MS tape. Additionally, absorbers were deployed to reduce unnecessary scattering from the receiver and the GND plane edges. 

\begin{figure}[tb!]
\includegraphics[width=0.95\linewidth]{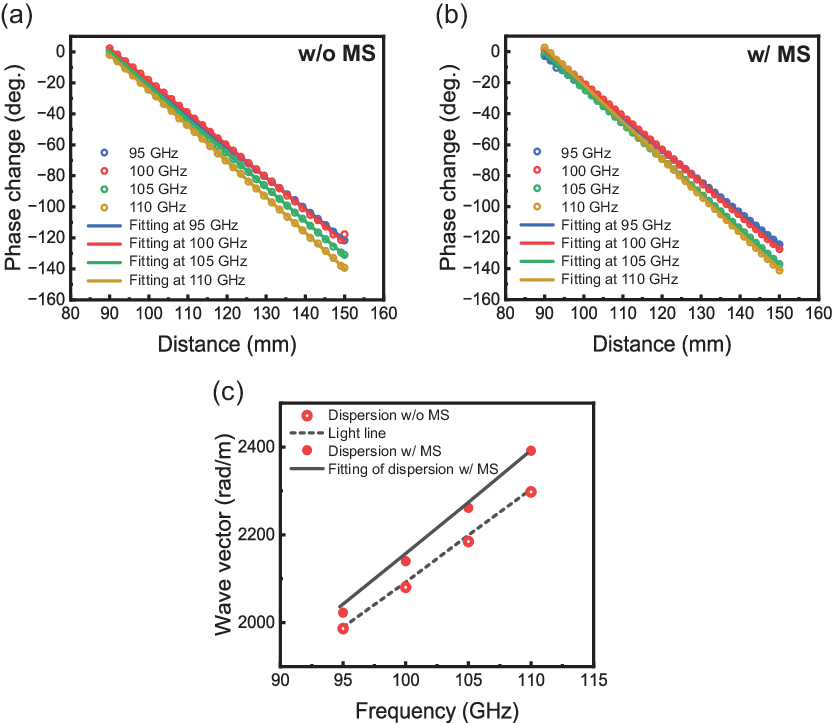}% Here is how to import EPS art
\caption{\label{figS:2} Phase change within our MS tape. (a) The phase change without the MS. (b) The phase change with the MS. (c) The corresponding wave vectors. }
\end{figure}

In Fig.\ \ref{fig:6}, we evaluated the received power (measured at the Rx horn) with the MS tape, while the corresponding phase variations are clarified in Fig.\ \ref{figS:2}. Note that Figs.\ \ref{figS:2}a and \ref{figS:2}b represent the phases extracted from simulations using the same model as in Fig.\ \ref{fig:2}c to avoid the standing wave issue shown in Fig.\ \ref{fig:3}. These results were then approximated by linear functions whose gradients are plotted in Fig.\ \ref{figS:2}c. In this figure, the frequency dispersion extracted from Fig.\ \ref{figS:2}a (i.e., the results without the MS) closely matched the light line. Similarly, the frequency dispersion extracted from Fig.\ \ref{figS:2}b (i.e., the results with the MS) was approximated by a linear function. According to these two straight lines, the relative differences in the wave vector between the cases with and without the MS tape at 95, 100, 105, and 110 GHz were 1.81, 2.88, 3.52, and 4.04 \%, respectively, which are sufficiently small for ordinary wireless communications. 

\begin{figure}[tb!]
\includegraphics[width=0.95\linewidth]{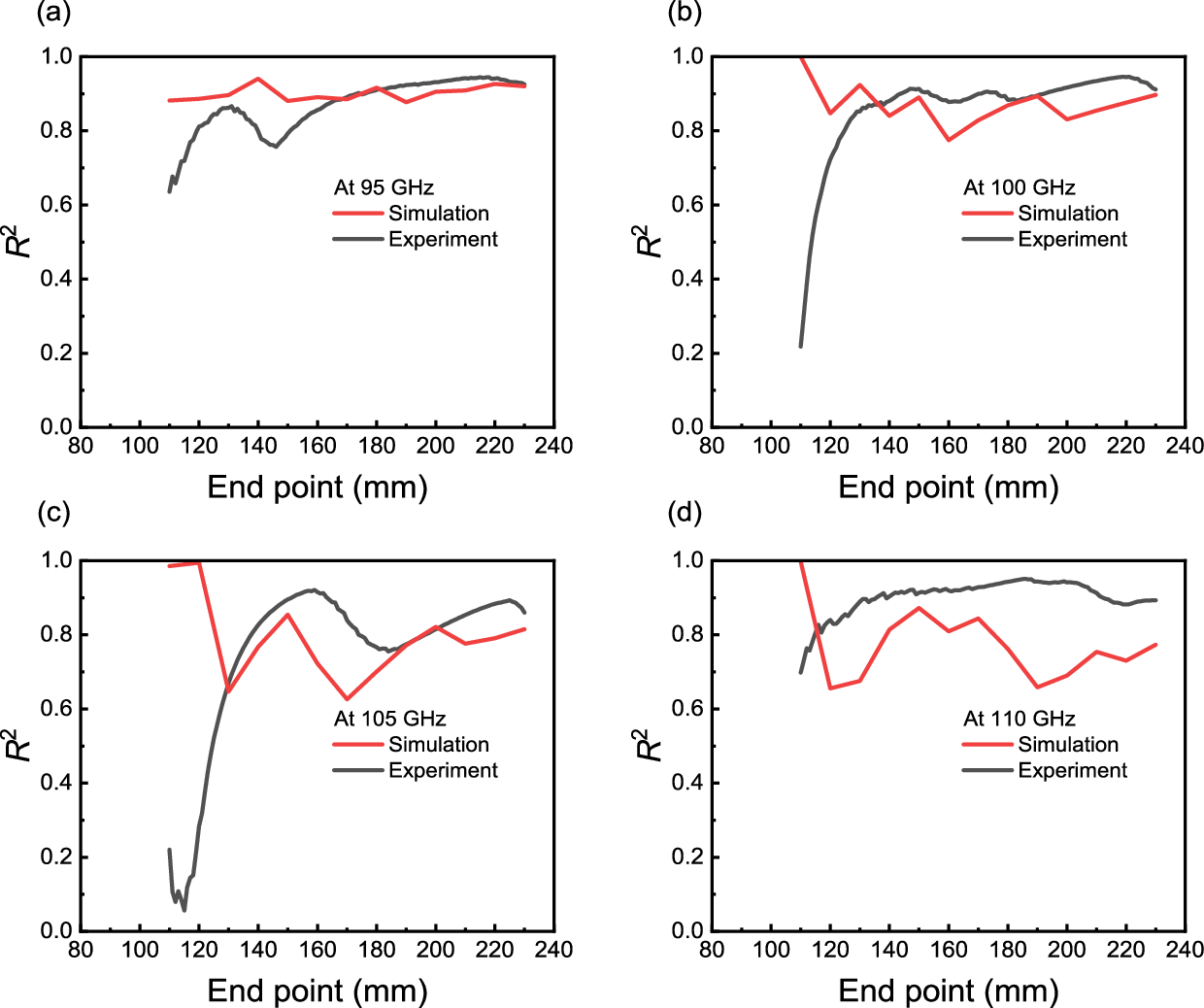}% Here is how to import EPS art
\caption{\label{figS:3} $R^2$ with various fitting ranges. (a-d) The results at (a) 95 GHz, (b) 100 GHz, (c) 105 GHz, and (d) 110 GHz. }
\end{figure}

\begin{figure}[tb!]
\includegraphics[width=0.95\linewidth]{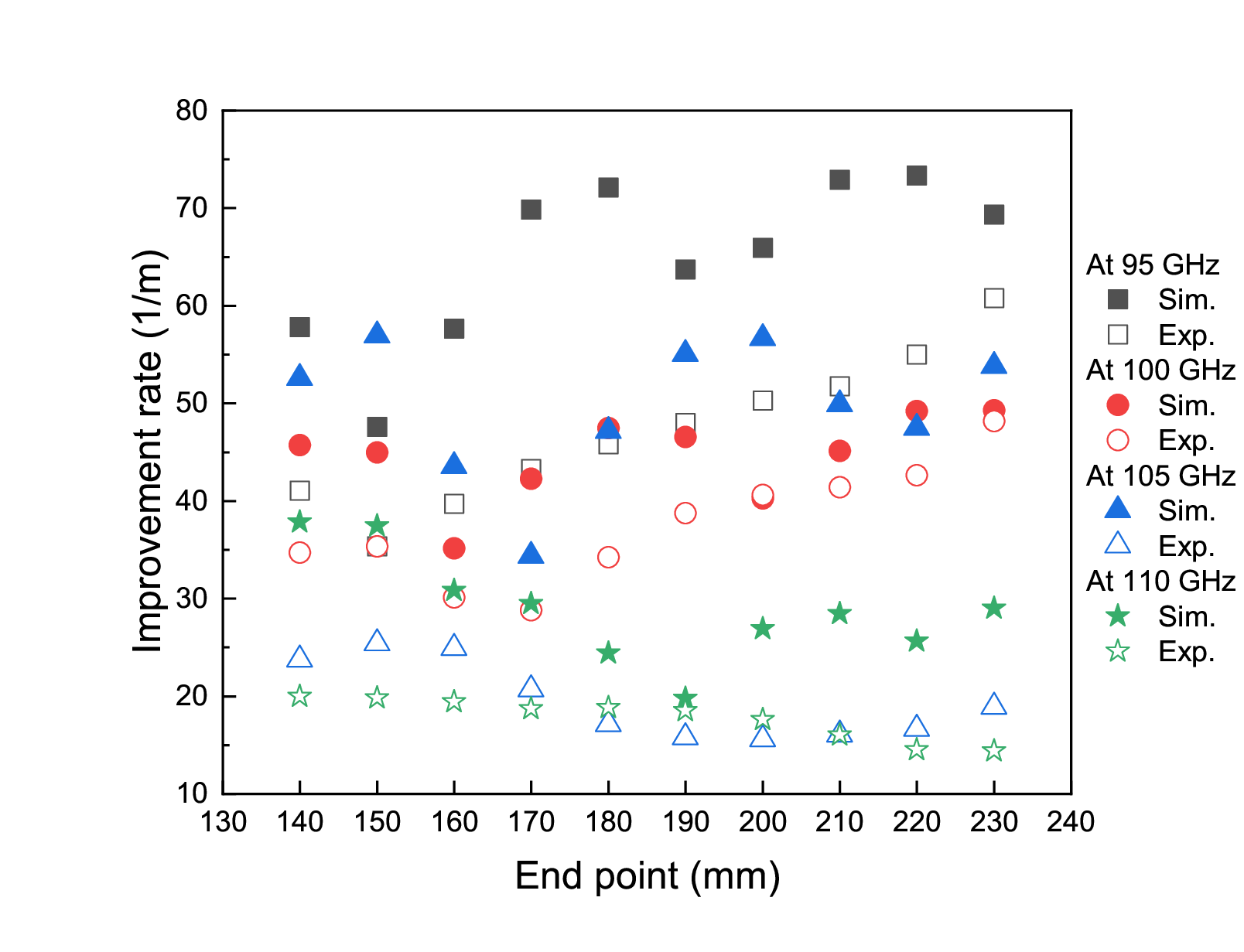}% Here is how to import EPS art
\caption{\label{figS:4} Rate of improvement factor $G$ with various end points. }
\end{figure}

Fig.\ \ref{fig:5} showed the improvement in received power level along with the corresponding linear fitting lines. However, the accuracy of these approximations depends on the fitting range, as clarified in Fig.\ \ref{figS:3}. This figure shows the coefficients of determination ($R^2$) obtained with the fixed starting point (90 mm from the Tx aperture) and various end points. According to these results, the $R^2$ values tended to be relatively low if the end point of the fitting distance was close to the starting point. Furthermore, the $R^2$ values obtained in the measurements were slightly low near 230 mm, presumably due to the influence of the MS tape edge. To avoid this influence, we set the fitting range from 90 mm to 220 mm in Fig.\ \ref{fig:5}. In Fig.\ \ref{figS:4}, we show how the fitting range influences the improvement rate. This figure also supports that the improvement rate was nearly converged when the end point of the fitting range was set to 220 mm.
\end{document}